\documentclass[prl,twocolumn,10pt, aps,superscriptaddress]{revtex4-2}
\usepackage{graphicx} 
\usepackage{amsmath}
\usepackage{color}

\def\d{\mbox{d}}  
\def\E{\mbox{e}}  

\def\kvec{\mathbf{k}}
\def\evec{\mathbf{e}}
\def\qvec{\mathbf{q}}
\def\vvec{\mathbf{v}}

\def\Jvec{\mathbf{J}}
\def\Svec{\mathbf{S}}

        \definecolor{AAcolor}{rgb}{0.7,0.1,0.4}

		\newcommand{\q}[1]{Eq.\ (\ref{#1})}
		\newcommand{\qq}[2]{Eqs.\ (\ref{#1})-(\ref{#2})}
		
		\newcommand{\fig}[1]{Fig.\ \ref{#1}}

\newcommand{\sma}[1]{\scriptscriptstyle{#1}}

\newcommand{\mathsout}[1]
{\bgroup\mathchoice
  {\sbox0{$\displaystyle{#1}$}%
    \usebox0\hspace{-\wd0}%
    \rule[0.5\ht0-0.5\dp0-.5pt]{\wd0}{1pt}}%
  {\sbox0{$\textstyle{#1}$}%
    \usebox0\hspace{-\wd0}%
    \rule[0.5\ht0-0.5\dp0-.5pt]{\wd0}{1pt}}%
  {\sbox0{$\scriptstyle{#1}$}%
    \usebox0\hspace{-\wd0}%
    \rule[0.5\ht0-0.5\dp0-.5pt]{\wd0}{1pt}}%
  {\sbox0{$\scriptscriptstyle{#1}$}%
    \usebox0\hspace{-\wd0}%
    \rule[0.5\ht0-0.5\dp0-.5pt]{\wd0}{1pt}}%
\egroup}

\newcommand{\bk}{\boldsymbol{k}}

\newcommand{\bq}{\boldsymbol{q}}

\newcommand{\bJ}{\boldsymbol{J}}

\newcommand{\bS}{\boldsymbol{S}}

\newcommand{\bpm}{\begin{pmatrix}}
\newcommand{\epm}{\end{pmatrix}}

\newcommand{\bal}{\begin{align}}

\usepackage{bm}
\usepackage{multirow}
\usepackage{dcolumn}
\usepackage{xcolor}
\usepackage{comment}
\newcommand{\batio}{BaTiO$_3$}

\usepackage{soul}

\begin{document}

\title{Photogalvanic currents from first-principles real-time density-matrix dynamics}

\author{Junting Yu}
\affiliation{Department of Materials Science and Engineering, University of Wisconsin-Madison}
\author{Andrew Grieder}
\affiliation{Department of Materials Science and Engineering, University of Wisconsin-Madison}
\author{Jacopo Simoni}
\affiliation{Department of Materials Science and Engineering, University of Wisconsin-Madison}
\author{Rafi Ullah}
\affiliation{Department of Materials Science and Engineering, University of Wisconsin-Madison}
\author{Zihao Bai}
\affiliation{Department of Chemistry, University of Wisconsin-Madison}
\author{Ravishankar Sundararaman}
\affiliation{Department of Materials Science and Engineering, Rensselaer Polytechnic Institute}
\author{Aris Alexandradinata}
\email{aalexan6@ucsc.edu}
\affiliation{Department of Physics and Santa Cruz Materials Center, University of California, Santa Cruz}
\author{Yuan Ping}
\email{yping3@wisc.edu}
\affiliation{Department of Materials Science and Engineering, University of Wisconsin-Madison}
\affiliation{Department of Physics, University of Wisconsin-Madison}
\affiliation{Department of Chemistry, University of Wisconsin-Madison}
\date{\today}

\begin{abstract}
The photogalvanic effect  is  the generation of a second-order direct current by illumination of a non-centrosymmetric material. 
In this work, we develop a first-principles real-time density matrix (FPDMD) formalism enabling the calculations of the photogalvanic current in all time regimes: transient and steady. Unlike past \textit{ab-initio} studies which focused only on the photo-excitation process, our first-principles theory framework encodes all quantum scatterings (intra/interband relaxation and electron-hole recombination) mediated by bosons (photons and phonons), and is thus predictive of photogalvanic currents in realistic materials. In particular, for the linear photogalvanic effect, we find electron scatterings mediated by phonons contribute significantly to the shift current for prototypical piezoelectrics like BaTiO$_3$.  
For the circular photogalvanic effect, we develop a self-consistent theory of a steady injection current that incorporates realistic scattering mediated by phonons. 
Our formulation developed for photogalvanic current 
elucidates its connection with fundamental quantum-geometric quantities such as the Berry curvature and the quantum metric. A phonon-based explanation is proposed for the bipolar transient photogalvanic current observed by the THz emission spectroscopy.
\end{abstract}

\maketitle

The photogalvanic effect (PGE), also known as the bulk photovoltaic effect, is the light-induced generation of a direct current in a homogeneous, non-centrosymmetric medium, without inhomogeneous doping or externally applied fields \cite{belinicher1980photogalvanic,fridkin2001bulk}. Interest in the photogalvanic effect has resurged, owing to recently unveiled 
connections with the quantum geometry of electronic wave functions  in topological materials\cite{tan2016enhancement, de2017quantized,zhu2024anomalous} and superlattices. This has led to envisioned applications to polarization- and frequency-sensitive photodetectors, as well as efficient solar cells capable of harvesting light with photon energy below 1 eV~\cite{zhu2024anomalous}.
Other applications of the PGE include the detection of centrosymmetry breaking in chiral and ferroelectric materials ~\cite{ma2017direct,le2021topology,li2021enhanced,zenkevich2014giant,feng2025high}, the detection of spin-splitting due to Rashba/Dresselhaus spin-orbit interactions~\cite{mcconnell2025coexisting,grieder2025relation}, as well as tracking the ultrafast dynamics of electron distributions and order parameters 
\cite{pettine2023ultrafast, sotome2019spectral, okamura2022photovoltaic, subedi2025colossal}. 

\begin{figure*}[ht]
    \centering
    \includegraphics[scale=0.55]{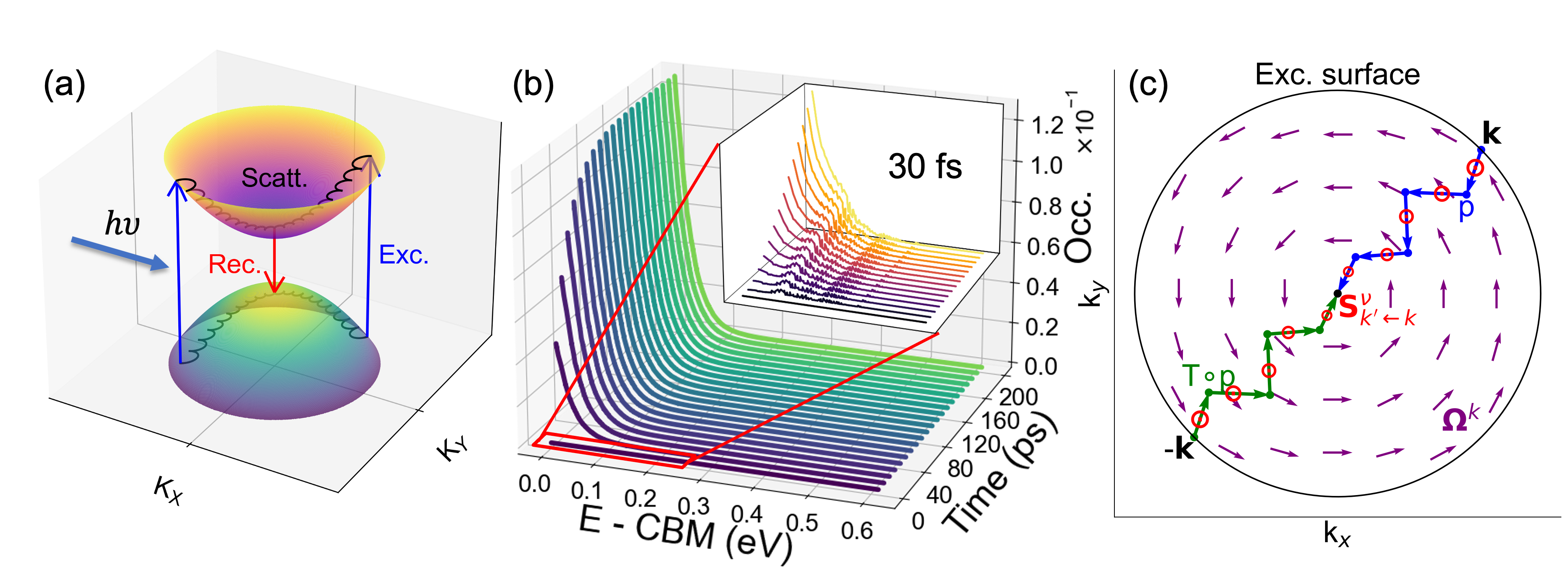}
    \caption{Kinetic process and electron occupation under an external light illumination. (a) Kinetic processes including light excitation, scattering and recombination. (b) Excited carrier occupation changes with time in conduction bands within 200 ps;
    the inset zooms in the first 30 fs.  The excitation peak is formed within 10 fs, but conduction electrons are then quickly scattered to conduction band edge. Within $25$ fs (the energy relaxation time), the precursor of a quasi-Fermi-Dirac distribution forms, with a quasi temperature 315K. 
    (c) For the conduction band of BaTiO$_3$,  the intra-band Berry curvature $\mathbf{\Omega}^\kvec$ is represented as a  purple vector field enclosed by the excitation surface at $\hbar\omega-E_g=0.32$ eV.  Under  electron-phonon scattering, the blue arrows ($p$) illustrate one representative electron trajectory from $\kvec$ to band bottom, while the green arrows ($T\circ p$) is the trajectory that is the time-reverse of $p$. For a small-momentum scattering event from $\kvec$ to $\kvec'$, the electron is displaced in real space by the phononic shift vector $\Svec^\nu_{\kvec'\leftarrow\kvec}=\mathbf{\Omega}^{\bar{\kvec}}\times (\kvec'-\kvec)$ with $\bar{\kvec}=(\kvec+\kvec')/2$. This  cross product of a purple arrow and a blue/green arrow is  represented as  
    a red circle and points in the z direction, which is the polar axis of BaTiO$_3$. The shift contributions from $p$ and $T\circ p$ add up instead of canceling out, because under time-reversal,  $\mathbf{\Omega}^{-\bar{\kvec}}=-\mathbf{\Omega}^{\bar{\kvec}}$ and $(\kvec-\kvec')=(-\kvec+\kvec')$, leaving the shift vector invariant.}
    \label{fig:1}%
\end{figure*}

Here we develop a first-principles real-time density matrix dynamics (FPDMD) formalism for the PGE, based on quantum master equations (QME) which encode all quantum kinetic scatterings  mediated by bosons (photons and phonons), including not just the well-studied photo-excitation process but also the intra- and interband relaxations and the electron-hole recombinations illustrated in Fig.~\ref{fig:1}(a). Our theory is thus predictive of the photogalvanic current in realistic materials where electron-phonon scattering cannot be ignored, and promises to resolve controversy and confusion about the underlying scattering mechanisms that contribute to the current. 
Contrary to the common misconception that the shift photogalvanic current (generated in nonmagnetic systems under linearly polarized light) is purely a photo-excitation effect, we demonstrate that phonon-mediated scattering will contribute significantly to the shift current in prototypical piezoelectrics. 
With first‑principles real-time density matrix
formalism, we go beyond prior model-Hamiltonian studies of PGE in topological materials~\cite{zhu2024anomalous}, and we demonstrate that phonon contributions to the shift current are ubiquitous across piezoelectrics. 
Next, we develop a self-consistent theory of the steady photogalvanic current under circular polarized light, with a state-dependent
relaxation time $\tau_{\bf{k}s}$ and the non-equilibrium steady electron distributions. This goes beyond previous perturbative calculations of the linear growth $\d J/\d t$ \cite{sipe2000second}, where a single relaxation time approximation $\tau_0$ does not self-consistently describe the
multiple scattering mechanism in realistic materials. 
Lastly, despite recent investigations of the transient photogalvanic
current through THz emission experiments~\cite{braun2016ultrafast,sotome2019spectral,sotome2019ultrafast,sotome2021terahertz} and numerical simulations~\cite{tan2024dynamics}, a rigorous explanatory theory remains to be
developed.
Our FPDMD opens a unique pathway into transient photogalvanic currents with femtosecond resolution. 
In particular, the experimental
observation~\cite{braun2016ultrafast} of a bipolar $J(t)$ is explained
as a dynamical cross-over from a photo-excitation-dominated
current to an electron-phonon-dominated current.

To set up these results, the total Hamiltonian for an electron-boson system
\begin{gather}
H_T = H^e_0 + H^b_0 + H' 
= \sum_i\varepsilon_i c^\dagger_i c_i + \sum_m\hbar\omega_m a^\dagger_m a_m + H'. \notag
\end{gather}
is a sum of three terms: (i) $H^e_0$ is the unperturbed single-particle electronic Hamiltonian with a Bloch basis computed from first-principles Kohn-Sham Density-Functional Theory (KSDFT) and  $c_i=c_{\kvec,s}$ annihilates an electron with wavevector $\kvec$ and band label $s$. (ii)
$H^b_0$ is the unperturbed boson Hamiltonian, and $a_m$ annihilates a boson (phonon or photon)  with mode index $m$. (iii) The interaction Hamiltonian is
\begin{equation}
H' = \sum_m\sum_{ij}\lambda^m_{ij}c_i^\dagger c_j \otimes (a_m + a^\dagger_{-m}),
\label{eqn:H_prime}
\end{equation}
where $\lambda^m_{ij}$ is the electron-boson interaction matrix element:
\begin{equation}
\lambda_{ij}^{m} = 
\begin{cases}
\delta_{\kvec_i\kvec_j+\textbf{q}} \sqrt{\frac{e^2\hbar}{2\epsilon_0\omega V}} \evec\cdot\vvec^{\kvec_i, \kvec_j}_{s_is_j},  m=(\textbf{e}, \textbf{q})\ \ \text{for photon} \\
\delta_{\kvec_i,\kvec_j+\qvec} \ g^{\nu\kvec_i\kvec_j}_{s_is_j}, m=(\nu, \bf{q})\ \ \text{for phonon}.
\end{cases} \notag
\end{equation}
Focusing first on the electron-photon interaction, we use velocity gauge,  
$\vvec^{\kvec_i,\kvec_j}_{s_is_j}=\langle \phi_{\kvec_i s_i}|e^{i\textbf{q.}\textbf{r}}\textbf{v}|\phi_{\kvec_js_j}\rangle$, 
with $\phi_{\kvec_i s_i}$ being single particle wavefunctions, $\vvec=\frac{i}{\hbar}[\mathbf{r}, H^e_0]$ being the velocity operator and $H^e_0$ including the non-local part of the atomic pseudopotential in KSDFT. We consider direct optical transitions in the dipole approximation at long wavelength limit $\textbf{q}\rightarrow 0$. 
If we assume the light source is monochromatic with a classical electromagnetic vector potential, the electron-photon interaction will be connected with vector potential via 
$\lambda^m_{ij}=\delta_{\kvec_i\kvec_j}(e/2\sqrt{N})\mathbf{A}_0\cdot\vvec^{\kvec_i}_{s_is_j}$
with $\mathbf{A}_0=\sqrt{2\hbar N/(\epsilon_0\omega V)}\evec$ the amplitude of light vector potential, $\evec$  the light polarization vector (which may be real for linearly-polarized light and must be complex for circularly-polarized light), and $N$ the light-source photon occupation number.
For the electron-phonon interaction, $\delta_{\kvec_i, \kvec_j+\qvec}$ is the momentum conservation constraint, and $g^{\nu\kvec_i\kvec_j}_{s_is_j}$ is the first-order electron-phonon matrix element (for a phonon eigenmode $\nu$) computed from density-functional perturbation theory~\cite{Baroni2001,Giustino2017,Note1} 
\footnotetext[1]{For more details see SI sec IV}. 

The QME describes the time evolution of the electron reduced density-matrix (RDM) accounting for photoexcitation (exc), electron-phonon scattering (ph), and electron-hole recombination (rec) processes~\cite{Xu2024-jj,Simoni2025-ty,Xu2021-xv}:
\begin{equation}
\frac{\d\rho}{\d t} = -\frac{i}{\hbar}[H^e_0, \ \rho] +  \frac{\d\rho}{\d t}|_{\mathrm{exc}} + \frac{\d\rho}{\d t}|_{\mathrm{ph}} + \frac{\d\rho}{\d t}|_{\mathrm{rec}}.
\label{qme1}
\end{equation}
While in principle time-dependent electron-electron scattering  can also influence RDM dynamics (discussed in our work on open quantum dynamics formulation~\cite{Simoni2025-ty} and ultrafast spin dynamics~\cite{Xu2024-jj}),  in practice it is only relevant at low temperature~\cite{Xu2021-xv, xu2024spin} or 
high-intensity light sources\cite{esipov1987temperature}, which we avoid in this work.
The charge current density is evaluated as $\Jvec(t)=-e\mathrm{Tr}(\rho(t)\vvec)=\sum_{n=0}^{+\infty}J(\omega)e^{in\omega t}$, with different Fourier components giving the direct current ($n=0$), second ($n=2$), and higher harmonic generation. It is useful to decompose $\Jvec=\Jvec^{d}+\Jvec^{\text{off}}$, such that the band-diagonal  current $\Jvec^{d}$ (band-off-diagonal current $\Jvec^{\text{off}}$) is contributed by the band-diagonal (resp. band-off-diagonal) matrix elements of $\rho$.
Other types of current can be obtained by evaluating the corresponding observable, e.g. the spin photocurrent density $\Jvec_{\bS}(t)=\mathrm{Tr}(\rho(t)(\Svec\vvec+\vvec \Svec)/2)$ ~\cite{xu2021pure,song2021spin}.

\begin{figure*}[ht]
    \centering
    \includegraphics[scale=0.3]{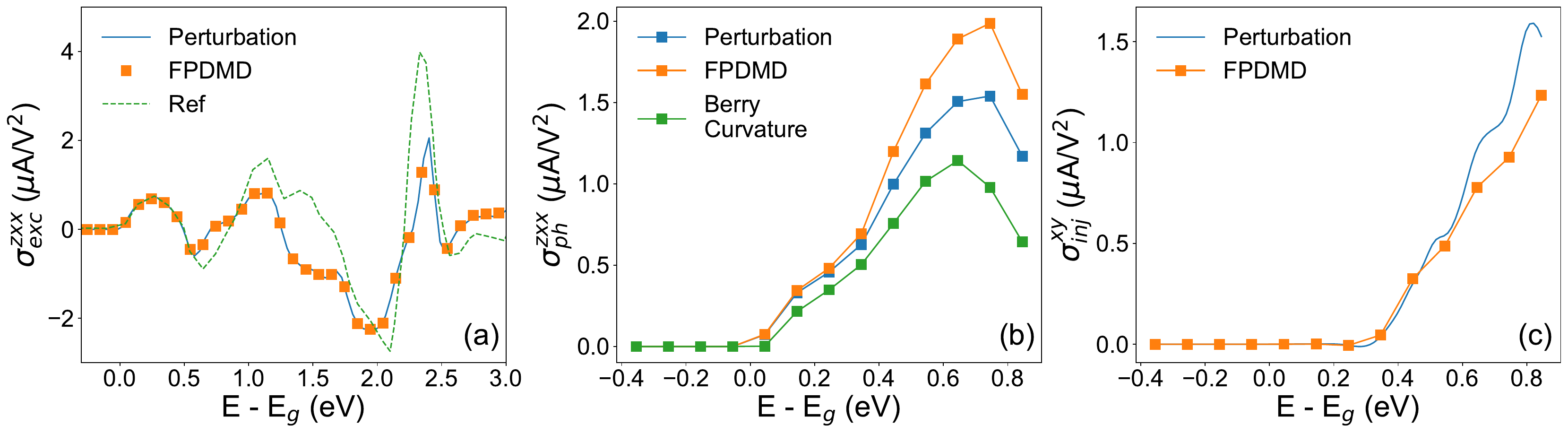}
    \caption{
    Different DC photocurrent conductivities of BaTiO$_3$. (a) Excitation shift current conductivity in $zxx$ direction. Real time  result is from evaluating the steady-state off-diagonal part of the density matrix, with eliminating other contributions except excitation current. Reference data is from Ref.\cite{fei2020shift}. (b) Phonon shift current conductivity from perturbation theory, FPDMD, and Berry curvature formula at RT in $zxx$ direction. FPDMD is obtained by eliminating other contributions except phonon shift current. Perturbation refers to results calculated by Eq.\eqref{eqn:shift_ph}. The Berry curvature results are obtained by replacing $\Svec^\nu_{\kvec's\leftarrow \kvec s}$ with $\qvec\times\mathbf{\Omega}^{\bar\kvec}_{ss}$, and only considering intra-band electron-phonon scattering in Eq.\ref{eqn:shift_ph}. (c) Injection current conductivity of BaTiO$_3$ from perturbation theory (Eq.~\ref{eqn:Jinj}) and FPDMD.  The perturbation result uses state-dependent electron-phonon relaxation rate (Eq.~\ref{statedepGamma}). 
    }
    \label{fig:2}%
\end{figure*}

Under the Born-Markov approximation,
the QME of electron-boson interaction has the following form \cite{breuer2002theory,rosati2014derivation},
\begin{gather} 
\frac{\d\rho_{\sma{12}}}{\d t}\bigg|_c {=} \sum_{\sma{345}}
(I-\rho)_{\sma{13}}\rho_{\sma{45}}P^c_{\sma{3245}} 
- \rho_{\sma{32}}(I-\rho)_{\sma{45}}P^c_{\sma{5431}}
{+}{(1{\leftrightarrow}2)^*} \nonumber \\
P^c_{1234} = \frac{\pi}{\hbar}\sum_{m\in c,\pm}N_m^{\mp} \Lambda^{\pm m\mp}_{13}\lambda^{\pm m*}_{24}, \label{qme2}
\end{gather}
for each scattering channel $c=\mathrm{exc,ph,rec}$; $\Lambda^{m\pm}_{12}=\lambda^m_{12}\delta(\varepsilon_1-\varepsilon_2\pm\hbar\omega_m)$,
$N_{m}^{\pm}=N_{m}+\frac{1}{2}\pm\frac{1}{2}$, and $N_{m}$ are boson occupations.

Our case study focuses on tetragonal BaTiO$_3$, a prototypical piezoelectric with a long history of non-linear optical measurements \cite{von1950ferroelectricity,cardona1965optical,koch1975bulk, ehsan2021first, dwij2022revisiting}.
%
Fig. \ref{fig:1} presents the carrier dynamics of this system from FPDMD: (b) shows that within 10 fs of turning on the light source, the electrons form a peak located around the excitation energy $\varepsilon_{\kvec c}=\varepsilon_{\kvec v}+\hbar\omega \approx 1.9$ eV. 
Within 30 fs, close to the energy relaxation time $\tau_{er}=25$ fs in Fig.~\ref{fig:4}, electrons are scattered to the conduction band edge, and the aforementioned excitation peak evolves to a cusp (Fig.~\ref{fig:1}(b) inset). 
Within several hundred ps,  a quasi-Fermi-Dirac distribution develops for electrons and holes respectively, with a common fitted temperature T = 315 K, which is close to the phonon temperature of 300 K.

Focusing first on the PGE under linearly-polarized light, we isolate the \textit{excitation shift current} $\bJ_{exc}^{sh}$ by implementing the relaxation time approximation (RTA) for other processes except excitation: $\d\rho/\d t|_{ph} +\d\rho/\d t|_{rec}= -(\rho-\rho^{eq})/\tau_0$ for electron-phonon scattering and recombination processes, which effectively eliminates their contributions to the shift current \cite{Note2}\footnotetext[2]{More details can be found in SI sec II.A}.
In the steady state, the band-off-diagonal current is identified as  $\bJ^{\text{off}}=\bJ_{exc}^{sh}$. 
The second-order photocurrent conductivity tensor is then calculated as $\sigma^{abc}_{exc} = J^{\text{off}}_a/E_b(\omega)E_c(-\omega)$, with $E_b(\omega)=i\omega A_b$ being the electric field. 
Fig.~\ref{fig:2}(a) shows our real-time-simulated $\bJ_{exc}^{sh}$  agrees with the perturbative formula \cite{von1981theory,sipe2000second},
validating our FPDMD method for calculating non-linear photocurrents.

\begin{figure}[h]
    \centering
    \includegraphics[scale=0.3]{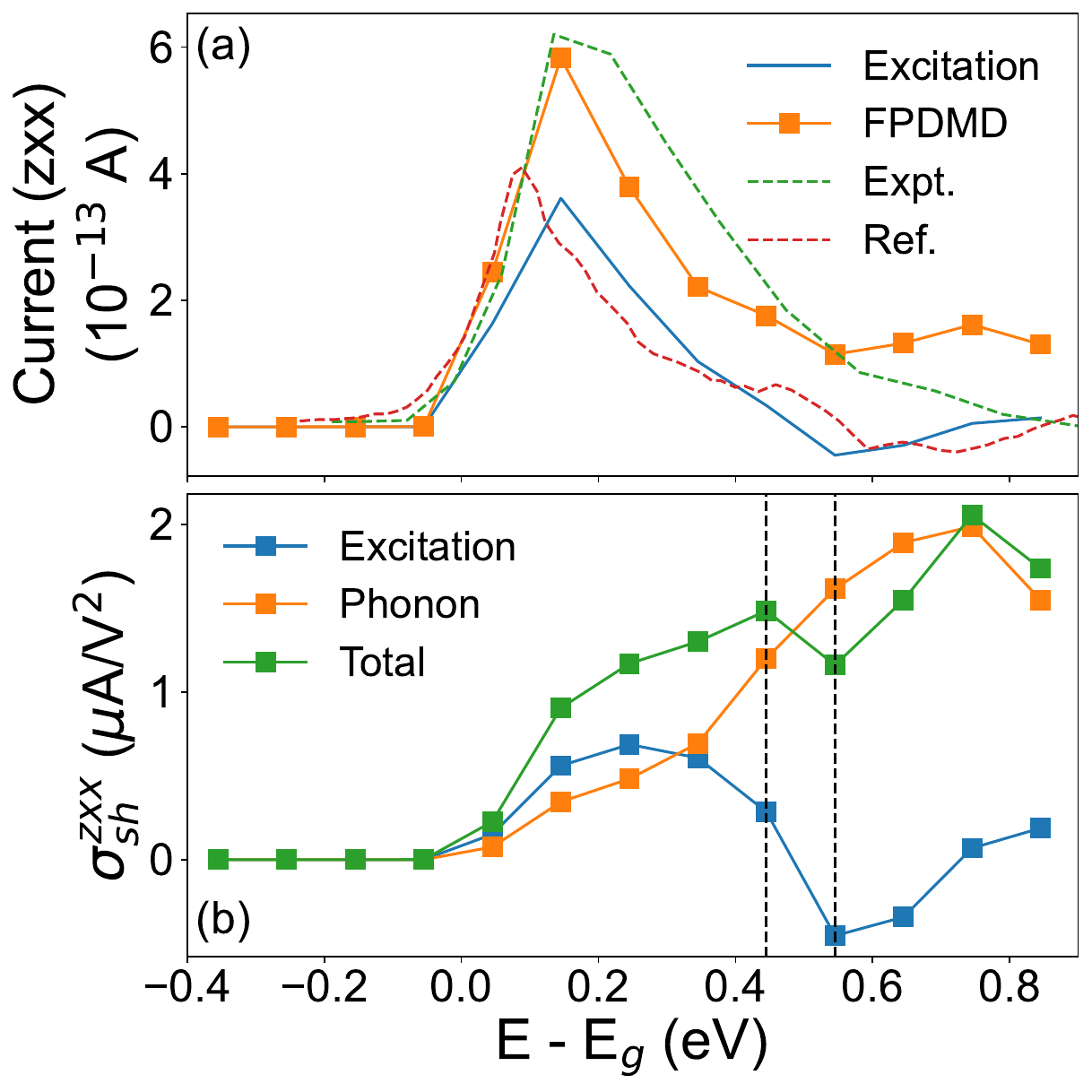}
    \caption{
    Photogalvanic current and conductivity of BaTiO$_3$ in $zxx$ direction. 
    (a) Comparison of  photocurrent (Eq.~\ref{dccurrent}) from real-time FPDMD, experiment result~\cite{koch1975bulk} and reference work \cite{dai2021phonon}. 
    (b) The total shift current conductivity and separate current components from FPDMD. 
    }
    \label{fig:3}%
\end{figure}

\begin{figure}%
    \centering    \includegraphics[scale=0.33]{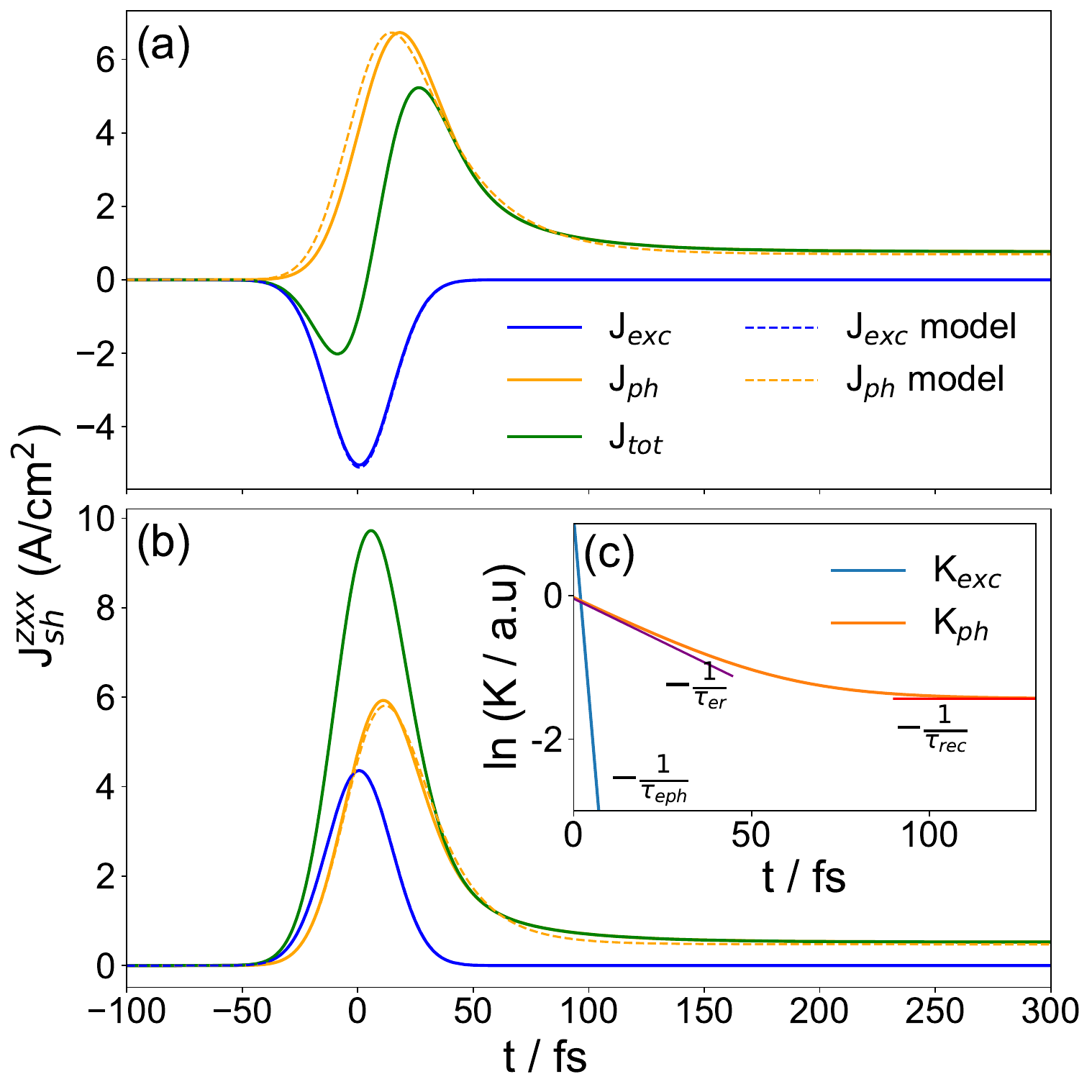}
    \caption{FPDMD simulation of $J(t)$ and their kernel-modeled results for BaTiO$_3$, using a Gaussian light pulse centered at $t=0$ with width 20 fs. 
    The modeled results are fitted against FPDMD. 
    Whether $J(t)=J_{tot}(t)$ is bipolar [as in panel (a) with photon energy at $0.545eV+E_g$] or unipolar [panel (b) with photon energy at $0.445eV+E_g$] is determined by whether the steady-state excitation and phonon shift currents  have opposite sign [the second dashed line in \fig{fig:3}(b)] or the same sign [the first dashed line in \fig{fig:3}(b)]. Panel (c) shows the fitting of real time dynamics to the convolution of $I(t)$ and $K(t)$ (kernel-models), where three different time scales are obtained, roughly in agreement with the electron-phonon scattering time $\tau_{eph}$, energy relaxation time $\tau_{er}$, and recombination time $\tau_{rec}$ of this system~\cite{Note7} (Details in SI Sec. IV). 
    }
    \label{fig:4}%
\end{figure}

Crucially, the phonon contribution to the shift current, which is missing in the aforementioned RTA, can only be calculated when explicit electron-phonon scattering is included in the QME. 
We extract this \textit{phonon shift current} $\bJ_{ph}^{sh}$ by subtracting the excitation shift current and recombination current from FPDMD total current. Fig.~\ref{fig:2}(b) shows  a semi-quantitative agreement between our FPDMD phonon-shift current and the perturbative, steady-state formula\cite{belinicher1982kinetic}:
\begin{gather}
\Jvec_{ph}^{sh;pert} 
= \tfrac{2e\pi}{\hbar V} \sum_{\sma{\kvec\kvec'ss'\nu}} \Svec^\nu_{\kvec's'\leftarrow\kvec s} |g^{\nu\kvec\kvec'}_{ss'}|^2\delta(\varepsilon_{\kvec s}-\varepsilon_{\kvec's'}+\hbar\omega_{\qvec\nu}) \nonumber \\
\times [(N_{\qvec\nu}+1)(1-f_{\kvec s})f_{\kvec's'} - N_{\qvec\nu}(1-f_{\kvec's'})f_{\kvec s}],
\label{eqn:shift_ph}
\end{gather}
where $\Svec^\nu_{\kvec's'\leftarrow\kvec s}$ is the phononic shift vector, $\qvec=\kvec-\kvec'$ is the phonon momentum, and $N_{\qvec\nu}=1/(\E^{\beta\hbar\omega_{\qvec\nu}}-1)$ is the thermal phonon occupation. Fig.~\ref{fig:3}(b) shows that $\bJ_{ph}^{sh}$ is comparable in magnitude to $\bJ_{exc}^{sh}$ at low light frequencies, but becomes dominant at higher light frequencies. 
This is because $\bJ_{exc}^{sh}$ is contributed only by one-electron states near the excitation $\bk$-surface (defined by $\varepsilon_{\kvec c}-\varepsilon_{\kvec v}=\hbar\omega$); at higher frequencies, the excitation $\bk$-surface encloses a larger volume of one-electron states whose mutual scatterings contribute to the phonon shift current~\cite{zhu2024anomalous}.


For intraband electron-phonon scatterings at small $\bq$, the phonon shift vector reduces to 
$\Svec^\nu_{\kvec's\leftarrow\kvec s}=\qvec\times\mathbf{\Omega}^{\bar{\kvec}}_{ss}+O(q^3)$\cite{zhu2024anomalous}, with $\bar{\kvec}=(\kvec+\kvec')/2$, and the intraband Berry curvature $\mathbf{\Omega^\kvec}=\nabla_\kvec\times\xi^\kvec$ being the curl of the  intraband Berry connection $\xi^\kvec_{ss}=i\langle u_{\kvec s}|\nabla_\kvec u_{\kvec s}\rangle$, with $u_{\kvec s}$ being the periodic part of single particle wavefunctions. 
By approximating the phonon shift vectors  in~\q{eqn:shift_ph} by their intraband Berry curvature formula, we obtain the approximate phonon shift current [green curve in Fig.~\ref{fig:2}(b)] that is semi-quantitatively consistent with the real time FPDMD current, indicating the prevalence of small-$\bq$ scattering and validating the quantum-geometric PGE theory \cite{zhu2024anomalous}. As explained under \fig{fig:1}(c), the circulation of the Berry curvature is intimately tied to the ferroelectricity in \batio; to flip  the circulation (from anticlockwise to clockwise) is to invert the direction of $J^{sh}_{ph}$ along the polar axis.

To compare our real-time, steady-state  PGE with the experimental data on BaTiO$_3$ under linear light, we  calculate the DC photocurrent as \cite{fei2020shift,cardona1965optical,Note3}\footnotetext[3]{More details can be found in SI sec VI}
\begin{equation}
\mathrm{Current} (abb) = \frac{\sigma_{abb}(\omega)}{\alpha_{bb}(\omega)}(1-R_{bb}(\omega))L\cdot |E_b(\omega)|^2, \label{dccurrent}
\end{equation}
with $\alpha_{bb}$ being the absorption coefficient, $R_{bb}$ the reflectivity, $L$ the sample width, and $E_b$ the electric field amplitude along the $b$ direction.
The comparison among excitation shift current, total current from FPDMD, and the experimental result is shown in Fig.~\ref{fig:3}(a). 
The excitation shift current is similar to the past perturbation theory result \cite{fei2020shift}.
It alone underestimates photocurrent compared with experiments, which is large between 0.0 and 0.6 eV above the band edge transition. 
Past work based on phonon-assisted ballistic current represents a different phonon-mediated mechanism from this work~\cite{dai2021phonon}. 
It enhances photocurrent between 0.0 and 0.2 eV, improved agreement with experiments\cite{dai2021phonon}. Our phonon contribution brings total photocurrent closer to experiments in a wider range of energy above the band edge transition. 
The improved agreement with experiments from both studies indicates that phonon-mediated scatterings plays a significant role in the PGE. 

Under circularly-polarized light excitation,  the injection current $\Jvec^{inj}$ is identified with the band-diagonal $\bJ^{d}$ calculated from FPDMD, and compared with the perturbative expression for the steady injection current density developed in this work due to photo-excitation:
\begin{gather}
\Jvec^{inj;pert} = i\frac{\pi e^3}{2\hbar V}|A_0|^2 \sum_\kvec \sum_{ss'} f_{ss'}^{\kvec}(\tau_{\kvec s}\vvec_{ss}^{\kvec} - \tau_{\kvec s'}\vvec_{s's'}^{\kvec}) \nonumber \\
\times(v^{b\kvec}_{s's}v^{c\kvec}_{ss'} - v^{c\kvec}_{s's}v^{b\kvec}_{ss'}) \delta(\varepsilon_{ss'}^{\kvec}-\hbar\omega),
\label{eqn:Jinj}
\end{gather}
which we derive in  SI III.A. Here, $A_0$ is the amplitude of the vector potential: $\mathbf{A}(t)=\frac{1}{2}A_0[\exp(-i\omega t)\evec+\exp(i\omega t)\evec^*]$, $\evec=\evec_b+i\evec_c$ is the complex polarization vector, $f^\kvec_{ss'}=f_{\kvec s}-f_{\kvec s'}$ is occupation difference between two bands, and $\tau_{\kvec s}$ 
is a state-dependent relaxation time.
Assuming that the carrier relaxation is dominated by electron-phonon scattering at room temperature, which is typical for most semiconductors \cite{Giustino2017,Xu2021-xv}, 
we relate the inverse relaxation time to the electron-phonon carrier lifetime \cite{giustino2017electron,Xu2024-jj},
\begin{equation}
\frac{1}{\tau_{\kvec s}} = \frac{2\pi}{\hbar}\sum_{\kvec's'}\sum_{\nu\pm}(N^\mp_{\qvec\nu}\pm f_{\kvec's'}) 
\delta(\varepsilon_{\kvec s}-\varepsilon_{\kvec's'}\pm\hbar\omega_{\qvec\nu})|g^{\nu\kvec\kvec'}_{ss'}|^2. \label{statedepGamma}
\end{equation}
Here, $f_{\kvec's'}$ is the steady electron distribution (photoexcited under light illumination, i.e. Fig.~\ref{fig:1}(b)), obtained from our FPDMD and  input to \qq{eqn:Jinj}{statedepGamma}, which represents a self-consistent solution. 
In previous, heuristic estimations of injection current\cite{gao2021intrinsic,de2017quantized,dai2023recent}, 
it is  a common  practice to replace the state-dependent $\tau_{\kvec s}$ by a state-independent relaxation time $\tau_0$, and $f_{\kvec s}$ by the equilibrium electron Fermi-Dirac distribution in the dark. The injection current conductivity can be calculated via $J^{inj}_a=2\sigma_{inj}^{ab}[\mathbf{E}(\omega)\times\mathbf{E}^*(-\omega)]_b$. 
We focus on the $\sigma_{inj}^{xy}$ component, the largest one for BaTiO$_3$.
Fig.~\ref{fig:2}(c) shows this conductivity in reasonably good agreement
between the FPDMD and the perturbative formula in \qq{eqn:Jinj}{statedepGamma}.

In the limit of a single conduction and a single valence band, the injection current conductivity in \qq{eqn:Jinj}{statedepGamma} can be reformulated in terms of the Berry curvature $\Omega^{b\kvec}$: 
\begin{equation}
\sigma^{inj}_{ab} = \frac{\pi e^3}{2\hbar V} \sum_{\kvec} \Omega^{b\kvec}_{11}(\tau_{\kvec1}v^{a\kvec}_{11} - \tau_{\kvec2}v^{a\kvec}_{22}) \delta(\varepsilon_{\kvec 12}-\hbar\omega).
\label{eqn:Jinj_berry}
\end{equation}
Here 1 and 2 represent conduction and valence band indices respectively. 
If $\tau_{\kvec1}$ is dominated by small-$\bq$, intraband electron-optical-phonon scattering, we may identify $|g^{\nu\kvec\kvec'}_{ss}|^2=\kappa_\nu (1-\sum_{ij}q_iq_j\zeta^{ij}_{\bar{\kvec}s})/q^2$ \cite{zhu2024anomalous}, with $\kappa_\nu$ being a non-zero constant for longitudinal optical (LO) phonons, and $\zeta^{ij}_{\kvec s}=  \mathrm{Re}\langle \nabla^i_{\kvec}u_{\kvec s}|\nabla^j_{\kvec}u_{\kvec s} \rangle -\xi^{i\kvec}_{ss}\xi^{j\kvec}_{ss}$ being the quantum metric \cite{provost1980riemannian}.
Thus, \qq{statedepGamma}{eqn:Jinj_berry} elegantly synthesize two fundamental quantum geometric characteristics which relate to two separate kinetic processes: the Berry curvature of one-electron states on the photo-excitation surface (as was emphasized heuristically in Ref.~\cite{de2017quantized}), and the quantum metric of one-electron states related by intraband electron-phonon scattering.

To simulate the transient photogalvanic current measured in recent ultrafast THz emission spectroscopy \cite{braun2016ultrafast,sotome2019spectral,kaplan2022twisted,tong2021ultraefficient,somma2014high}, we perform FPDMD for the transient shift current, by applying a linearly-polarized light pump $I(t)\propto\exp[-(t/\bar\tau)^2]$ centered at $t=0$ with width $\bar\tau=20$ fs and measuring the response current. 
The transient excitation, phonon and total shift current responses of BaTiO$_3$ from FPDMD are shown in \fig{fig:4} at two light frequencies. 
The $J_{exc}$, excitation-shift current, responses quickly to the pulse, with a time lag $\tau_{eph}=0.75$ fs that has the same magnitude as the electron-phonon scattering time. 
The $J_{ph}$, phonon-shift current, responds with a time lag $\tau_{er}\approx 25 fs$, close to the energy relaxation time, which is approximately the time that photoexcited electrons are scattered from the excitation surface to the band extrema \cite{lundstrom2002fundamentals}. The time lags $\tau_{eph/er}$ of $J_{exc/ph}$  are  respectively quantified by fitting each FPDMD-derived current to a time convolution of a Kernel function ($K_{exc}(t)$ or $K_{ph}(t)$) and the light intensity ($I(t)$). Here, each $K(t)$ is exponentially decaying with a decay constant equal to the corresponding time lag: $K(t)\propto \exp[-t/\tau]$, as detailed  in SI Sec. IV.
The two different time lags result in the total transient current being (a) bipolar when $J_{exc}$ and $J_{ph}$ have different signs, and (b) unipolar when $J_{exc}$ and $J_{ph}$ have the same sign. 
Our theory thus predicts a photon-frequency-dependent crossover between unipolar and bipolar $J(t)$, which possibly explains the puzzlingly bipolar $J(t)$ observed in THz emission spectroscopy of a number of systems\cite{braun2016ultrafast,sotome2019spectral,sotome2019ultrafast,sotome2021terahertz}. 

\footnotetext[7]{The three timescales from fitting kernel models to FPDMD are $\tau_{eph}=0.75$ fs, $\tau_{er}=25$ fs, and  $\tau_{rec}=23$ ps. Here $\tau_{eph}$ is close to $\langle\tau_{\kvec s}\rangle = 1.4$ fs, the averaged electron-phonon relaxation time over the excitation surface; $\tau_{er}$ is close to the time that excited electrons are scattered from excitation surface to band extrema, $\tau_{er}\approx\langle\tau_{\kvec s}(\varepsilon_{\kvec s}-\varepsilon_{CBM})/\hbar\omega_{LO}\rangle=21.2 fs$, where $\omega_{LO}=21.7 meV$ is the energy of the dominant LO mode; 
$\tau_{rec}$ is consistent with experimental observed recombination lifetime ranging  from $10ps$ to $100ps$ for different, impurity-varying samples of BaTiO$_3$. }

In summary, we develop a first-principles real-time density
matrix dynamics formalism for calculating the photogalvanic effect (PGE), enabling one to obtain different contributions from a unified theoretical framework. 
Taking BaTiO$_3$ as an example, we reproduce the excitation shift current and injection current that are consistent with perturbative methods. 
With explicit electron-phonon scattering, we find significant phonon shift current at excited states. 
Besides steady state DC, we are able to study the transient current that is directly accessible during the real time simulation, enabling us to understand photocurrent response in ultrafast THz spectroscopy experiments. 
Our work provides a general theoretical framework to study other contributions such as Coulomb scattering and higher-order scattering processes in the future \cite{tan2024dynamics,chan2021giant,lee2020ab}. 
Other non-linear optics such as second harmonic generation (SHG) or higher-order harmonic generation (HHG) can be extracted  straightforwardly from this theory \cite{lewenstein1994theory}, which is important for characterization of topological properties and novel quantum materials. 
Beyond the charge photocurrent, other observables can be readily evaluated, e.g. the spin photocurrent~\cite{xu2021pure, tiancheng2021} and orbital photocurrent~\cite{Mu2021-mi}, which recently sparked interest in the spin-optotronic and orbitronics~\cite{go2021orbitronics,Rappoport2023-qg} community.

\textbf{Acknowledgments} 
We thank Junqing Xu, Elio J Koenig, and Liang Tan for very helpful discussions. This work is primarily supported by the Computational Chemical Sciences program within the Office of Science of the DOE under Grant No. DE-SC0023301. Part of the work is supported by  the Air Force Office of Scientific Research under Award No. FA9550-21-1-0087. Calculations were carried out at the National Energy Research Scientific Computing Center (NERSC), a U.S. Department of Energy Office of Science User Facility operated under Contract No. DEAC02-05CH11231.

\bibliographystyle{apsrev4-2}
\bibliography{references.bib}
\end{document}